# Roll Angle Adjustment Dims Starlink Satellites

Anthony Mallama and Jay Respler

2023 March 1

Contact: anthony.mallama@gmail.com

Abstract

The brightness of Starlink satellites during orbit parking and orbit raising decreased significantly in 2020 when the operator modified their orientation. The mean apparent magnitude before the change was 3.90 +/- 0.18, while afterward it was 5.69 +/- 0.06. When magnitudes are adjusted to a standard distance of 1,000 km the means are 4.86 +/- 0.16 and 7.31 +/- 0.05. The difference at the standard distance indicates that spacecraft with adjusted roll angles are 90% fainter on average than the earlier ones.

1. Introduction

The operator of Starlink satellites has tried several methods for mitigating their negative impact on astronomical research and on aesthetic appreciation of the night sky. In order to reduce luminosity during parking and orbit raising the operator adjusted the attitudes of the spacecrafts (SpaceX, 2020). The new roll angle placed the Sun in the plane of the satellite body. This configuration, known as *knife-edge*, was intended to reduce brightness by decreasing the surface area of the satellite that reflects sunlight.

This paper addresses the effectiveness of the knife-edge configuration in dimming Starlink satellites. Section 2 describes observations that were recorded before and after the adjustment. Section 3 examines brightness statistics in relation to attitude and also reports the satellites' illumination phase functions. Section 4 gives the conclusions and discusses the results in the broader context of satellite brightness.

2. Observations and data description

Visual magnitudes for Starlink satellites have been recorded by author Respler (40.330$^o$ N, 74.445$^o$ W) since these spacecraft were first launched in 2019. The apparent magnitude is measured by comparison to reference stars in the same optical field of view. Proximity between the satellite and the stars accounts for variations in sky transparency and sky brightness. The method is described in more detail by Mallama (2022).

Observations posted on the SeeSat email archive at http://www.satobs.org/seesat/index.html were searched to find reports on Starlink satellites in their parking orbits and during orbit raising, that is, before the spacecrafts had reached their operation altitudes. The information extracted for this study included the satellite catalog number, observational values (apparent magnitude, year, month, day and hour) and satellite data (height above the Earth, range and phase angle).

The range (distance between the observer and the satellite) is used to compute the brightness of satellites at a standard distance of 1,000 km by applying the inverse square law of light. The phase angle is the arc distance measured at a satellite between the directions to the Sun and to the observer. The catalog number was used to separate the satellites according to the three models of Generation 1 Starlink spacecrafts: Original Design, VisorSat and Post-VisorSat. The ranges, phase angles and satellite models are discussed more fully in the next section.

3. Brightness analysis and results

The Starlink operator has made several attempts to ameliorate the impact of their satellites on astronomy. The knife-edge configuration described in Section 1 was one such effort. Another was adding a sunshade to the so-called VisorSat model of Starlink (SpaceX, 2020). This change occurred at about the same time as the roll angle adjustment and it successfully dimmed these satellites when observed at the operational altitude (Mallama and Respler, 2022). However, the operator stated that the shade was ineffective for reducing brightness during parking orbit and orbit raising. In any case, the sunshade was later omitted from Starlink satellites (Post-VisorSat model) for operational reasons. The three models are differentiated in the following analysis.

The brightness of low altitude Starlink satellites is plotted versus the year of observation in Figure 1. The time of the change to knife-edge configuration and the timeframes of the three Starlink models are also delineated. Significant dimming of the apparent magnitude and of that adjusted to 1,000 km is evident beginning in the year 2020. The similar levels of brightness during the eras of VisorSat and Post-VisorSat indicate that the sunshade did not affect brightness of these satellites at low altitudes. That finding is in agreement with the operator's statement about parking orbits and orbit raising. Thus, the observed fading must have been due to the knife-edge configuration.

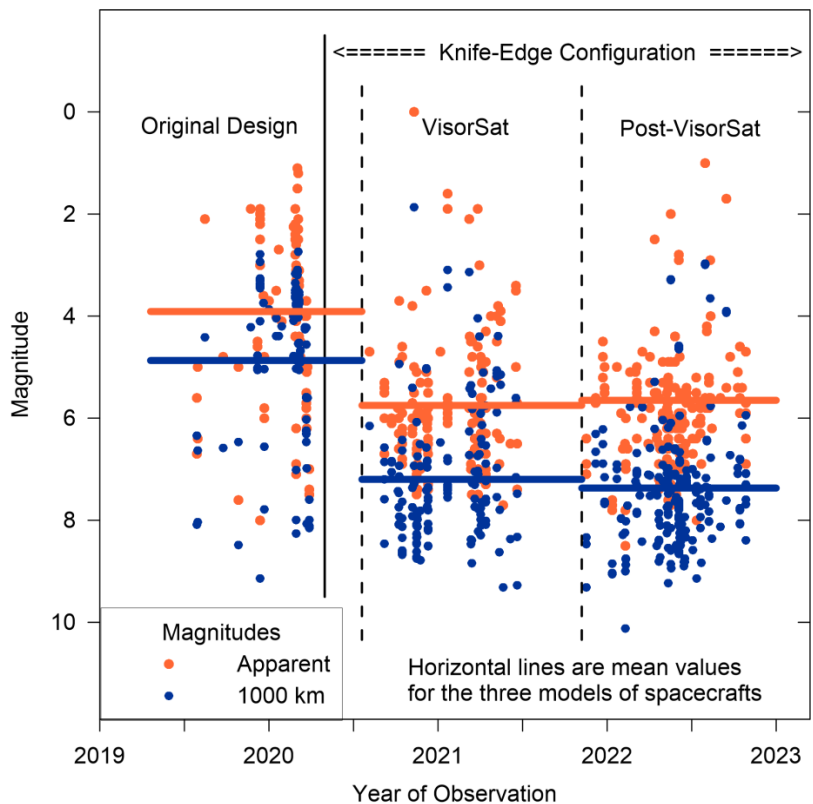

*Figure 1. Individual magnitudes and their averages as functions of year are illustrated. The timeframes of knife-edge configuration and of the three different Starlink satellites models are also shown.*

The brightness statistics of low altitude satellites listed in Table 1 reveal that dimming due to knife-edge orientation was 1.79 (that is, 5.69 minus 3.90) for apparent magnitudes and 2.44 (7.30 minus 4.86) for magnitudes adjust to 1,000 km. Thus, satellites in knife-edge configuration were only 19% and 10% (respectively) as bright as the others. The differing result for the two types of magnitudes is due to an observational selection effect. Specifically, the mean range for observations of knife-edge configured satellites was 301 km, while for the others it was 417 km. So, the apparent luminosity of knife-edge spacecraft was enhanced due to their smaller distances from the observer. Therefore, the value of 10% (taken at a standard distance) is a truer measure for the brightness of knife-edge oriented spacecrafts.

Table 1. Brightness statistics

```
            ---- Low Altitude ----    --- Operational Alt. ---    ------ Difference -----
               -- knife-edge --                                       -- knife-edge --
          OD      VS     PV    VS+PV    OD     VS     PV    VS+PV    OD     VS     PV    VS+PV
Apparent Magnitudes
Mean     3.90    5.75   5.65   5.69    5.31   6.30   5.69   6.02   -1.41  -0.54  -0.04  -0.33
StD      1.79    1.24   1.14   1.18    0.71   0.93   0.65   0.87
SDM      0.18    0.10   0.07   0.06    0.09   0.07   0.05   0.05

1000 km Magnitudes
Mean     4.86    7.20   7.37   7.30    5.93   7.22   6.71   6.99   -1.07  -0.03   0.66   0.31
StD      1.57    1.63   1.27   1.07    0.63   1.06   0.70   0.90
SDM      0.16    0.10   0.06   0.05    0.08   0.07   0.05   0.05

OD = Original Design, VS = VisorSat, PV = Post-VisorSat
StD = Standard deviation, SDM = Standard deviation of the mean
```

Additional insight on dimming can be obtained by comparing the low altitude brightness measures with those taken at operational altitudes. For that purpose, observations recorded by the same observer (Respler) have been taken from the study of operational altitude satellite by Mallama and Respler (2022). The brightness statistics for those higher altitude spacecrafts and the differences with low altitude satellites are also listed in Table 1.

For the Original Design satellites, which were not in knife-edge configuration, the low altitude satellites are much more luminous than those at operational altitudes. The mean values are brighter by 1.41 for the apparent magnitudes and 1.07 for the 1,000 km magnitudes. However, the low altitude spacecraft in knife-edge configuration are brighter by only 0.33 in apparent magnitude and are actually dimmer by 0.31 for magnitudes adjusted to 1,000 km. This is another indication of the effectiveness of the knife-edge configuration.

The illumination phase functions for low altitude satellites in Figure 2 show magnitudes adjusted to a standard distance of 1,000 km plotted versus phase angle. At large phase angles the Sun and satellite are in nearly the same direction as seen by the observer, and the spacecraft is illuminated from behind. At small angles the Sun and satellites are in opposite directions, and the side facing the observer is lit.

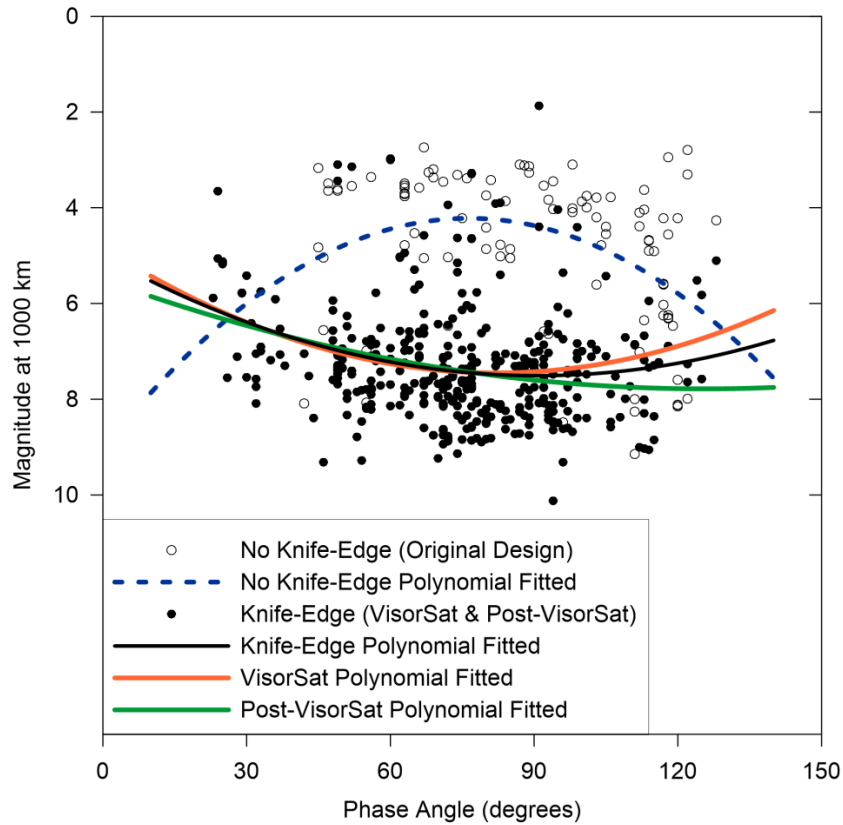

*Figure 2. Illumination phase functions for Knife-Edge (VisorSat and Post-VisorSat) and No Knife-Edge (Original Design). Observations are shown along with their polynomial fits. Fitted functions are also shown for VisorSat-only and Post-VisorSat-only magnitudes.*

The best fitting quadratic polynomial is upward curving for magnitudes of satellites in knife-edge configuration and downward curving for those not in knife-edge configuration. Fits to the knife-edge magnitudes indicate that these satellites are much fainter than the others near mid-range phase angle values which typically occur when a satellite is high in the sky. Meanwhile, the knife-edge and no-knife-edge fits cross at large and small phase angles which happen when a satellite is low in the sky and either toward or away from the Sun. Thus, the knife-edge configuration appears to be more effective when satellites are nearer zenith.

4. Conclusions and discussion

This study demonstrated how the brightness of Starlink spacecraft in orbit parking and orbit raising changed after their roll angles were adjusted so that the Sun was in the plane of the satellite body. Magnitudes transformed to a standard distance indicate that the roll adjustment reduced satellite luminosity by 90%.

The effect of spacecraft orientation on brightness demonstrated here for Starlink satellites is also seen in observations of the BlueWalker 3 satellite. Mallama et al. (2023) showed that that spacecraft was much fainter at times when it would have been tilted in order to increase the amount of sunlight falling on the solar array.

Satellite constellations are beginning to interfere with the work of astronomers making telescopic observations. Their impact on large research facilities such the Rubin Observatory has been addressed by Tyson et al. (2020). They are also a distraction for naturalists and others who appreciate the beauty of starry skies.

The International Astronomical Union has established a Centre for the Protection of Dark and Quiet Skies from Satellite Constellation Interference. The IAU-CPS maintains on an online presence where astronomers and satellite operators are able to meet virtually. These two communities can use the findings reported in this paper and other photometric studies of satellites (for example, Mroz et al. 2022 and Halferty et al. 2022) to redress the adverse impact of satellite constellations on astronomy.